# MgH$_2$ nanoparticles confined in reduced graphene oxide pillared with organosilica: a novel type of hydrogen storage material

*Feng Yan, Estela Moreton Alfonsín, Peter Ngene, Sytze de Graaf, Oreste De Luca, Huatang Cao, Konstantinos Spyrou, Liqiang Lu, Eleni Thomou, Yutao Pei, Bart J. Kooi, Dimitrios P. Gournis, Petra E. de Jongh, and Petra Rudolf\**

Dr F. Yan, E. M. Alfonsín, Dr S. de Graaf, Dr O. De Luca, Dr E. Thomou, Prof. B. J. Kooi, and Prof. P. Rudolf
Zernike Institute for Advanced Materials, University of Groningen, Nijenborgh 4, 9747 AG Groningen, the Netherlands.
E-mail: p.rudolf@rug.nl

Prof. P. Ngene, and Prof. P. E. de Jongh
Materials Chemistry and Catalysis, Debye Institute for Nanomaterials Science, Utrecht University, Universiteitsweg 99, 3584 CG Utrecht, the Netherlands.

Dr H. Cao, Dr L. Lu, and Prof. Y. Pei
Engineering and Technology Institute Groningen, University of Groningen, Nijenborgh 4, 9747AG Groningen, the Netherlands.

Dr K. Spyrou, Dr E. Thomou, and Prof. D. P. Gournis
Department of Materials Science and Engineering, University of Ioannina, 45110 Ioannina, Greece.

**Abstract**

Hydrogen is a promising energy carrier that can push forward the energy transition because of its high energy density (142 MJ kg$^{-1}$), variety of potential sources, low weight and low environmental impact, but its storage for automotive applications remains a formidable challenge. $MgH_2$, with its high gravimetric and volumetric density, presents a compelling platform for hydrogen storage; however, its utilization is hindered by the sluggish kinetics of hydrogen uptake/release and high temperature operation. Herein we show that a novel layered heterostructure of reduced graphene oxide and organosilica with high specific surface area and narrow pore size distribution can serve as a scaffold to host $MgH_2$ nanoparticles with a narrow diameter distribution around ~2.5 nm and superior hydrogen storage properties to bulk $MgH_2$. Desorption studies showed that hydrogen release starts at 50 °C, with a maximum at 348 °C and kinetics dependent on particle size. Reversibility tests demonstrated that the dehydrogenation kinetics and re-hydrogenation capacity of the system remains stable at 1.62 wt.% over four cycles at 200 °C. Our results prove that $MgH_2$ confinement in a nanoporous scaffold is an efficient way to constrain the size of the hydride particles, avoid aggregation and improve kinetics for hydrogen release and recharging.

# 1. Introduction

Today, the major source of energy are fossil fuels, the burning of which leads to global warming and climate change.[1,2] Hydrogen has been proposed as an alternative fuel already for several decades, since it represents an ideal zero-carbon energy carrier with a higher delivered energy-per-mass ratio (120 kJ/g) than conventional fuels such as petroleum (43.6 kJ/g) or coal (39.3 kJ/g).[3] Also, no $CO_2$ or $NO_x$ are produced in its combustion: the only byproduct is water vapor.[4] However, storage for automotive use remains a formidable scientific challenge due to the demanding high gravimetric and volumetric capacity required to satisfy competitive refueling needs.[5,6,7] Magnesium hydride, $MgH_2$, represents a particularly attractive platform as a solid-state hydrogen storage material owing to its high theoretical storage capacity (7.6 wt.% and 110 g/L) and outstanding reversibility that can potentially fulfill the requirements for fueling cars.[8,9] Moreover, magnesium is the eighth most abundant element in the Earth's mantle, and can therefore be considered a sustainable building block for a hydrogen storage material.[10]

Magnesium hydride still poses obstacles to the practical application since its thermodynamic stability with an enthalpy of approximately -75 kJ/mol leads to high operation temperatures.[11] In addition, the sluggish hydrogen absorption-desorption kinetics results in long times for hydrogenation and dehydrogenation.[12] The inherently low thermal conductivity[13] (2-8 W/(m·K)) also represents a difficulty for efficient use in this context. Nanostructuring of $MgH_2$ is a most effective strategy to lower the kinetic barrier,[14] since the large surface-to-volume ratio of the particles shortens the distance hydrogen atoms have to diffuse over.[15,16,17] Theoretical calculations and experimental studies have demonstrated that $MgH_2$ particle sizes of less than 5 nm lead to a significant enhancement of hydrogen adsorption/desorption kinetics.[18,19,20] However, $MgH_2$ nanoparticles have the tendency to minimize their surface energy by agglomeration when the temperature is high enough and this deteriorates the kinetic properties, leading to the slower cycling.[21,22] Confinement of $MgH_2$ particles in nanoporous supports has been demonstrated to be an effective way to improve hydrogen desorption properties, since the particle size can be easily controlled by modifying the pore size of the scaffolds, and the direct inter-particle contact is avoided, which can further prevent particle agglomeration.[23,24,25] Furthermore, the application of lightweight materials of high thermal conductivity can additionally counterbalance the loss of capacity and enhance the sluggish hydrogen sorption kinetics.[26,27]

Graphene, a one atomic layer thick two-dimensional material, has attracted wide attention in hydrogen storage because of its unique 2D structure, lightweight, and outstanding thermal conductivity (5300 W m$^{-1}$ K$^{-1}$).[28,29,30] In addition, graphene also acts as a catalyst for hydrogen dissociation/recombination.[31] Given these advantages, graphene has been combined with magnesium hydride through different methods, such as ball milling,[31] wrapping,[32,33] assembling,[26] confinement[9], *etc.* However, $MgH_2$ particles with a size less than 5 nm combined with graphene have not been achieved so far.

Inorganic-organic hybrids with organosilica building blocks prepared by surfactant directed sol-gel reaction of bridged organosilane precursors represent a new class of mesoporous materials.[34,35] In these hybrids, the organosilica functional groups can be integrated into the pore walls *via* the appropriate selection of organic precursors, and the pore size can be tuned through the selection of the surfactant.[36] Because of their high surface area, controllable mesoscale porous structure, light-harvesting properties and the possibility to achieve a high thermal stability by selecting the appropriate bridge group, these mesoporous organosilicas are promising for hydrogen storage.[37,38] Kalantzopoulos *et al.*[39], investigated both theoretically and experimentally phenylene-bridged organosilica in this context, and demonstrated that the pore size distribution appears to be the predominant factor for hydrogen storage and that reversible hydrogen adsorption capacities up to 2.1 wt.% can be realized at 6 MPa and 77 K.

In this work, we combine the advantages of graphene, or more precisely of reduced graphene oxide, rGO, and organosilica building blocks in a novel pillared heterostructure, which we synthesized by surfactant-directed sol-gel reaction of organosilica precursors in the interlayer space of graphene oxide, followed by removal of the soft template by pyrolysis. In this material, rGO assures thermal conductivity and a catalytic effect on $MgH_2$, while the organosilica pillars serve to confine the nanometer-sized $MgH_2$ particles and prevent them from aggregating. As a result, $MgH_2$ crystals with an average particle size of ~2.5 nm were grown in this heterostructure and showed outstanding hydrogen desorption properties. In addition, *in-situ* X-ray photoelectron spectroscopy was employed to elucidate structural changes during the dehydrogenation process and to gain insight into the dehydrogenation mechanism.

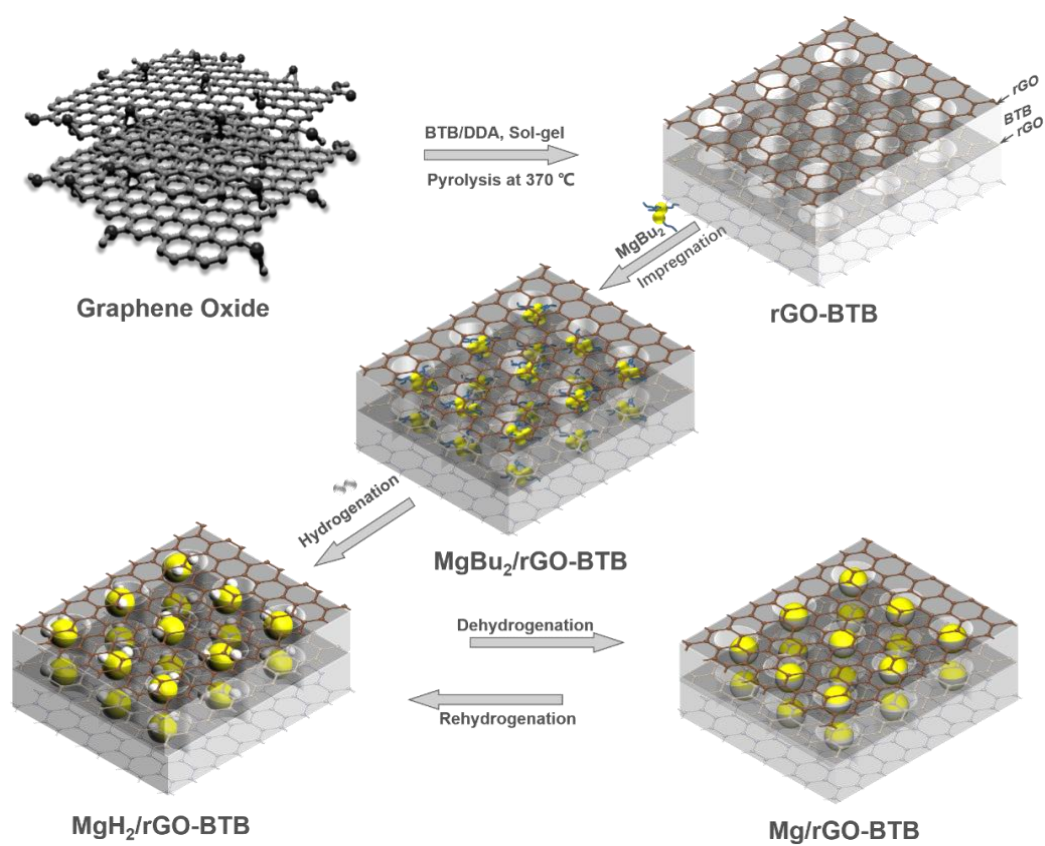

**Scheme 1. Schematic illustration of the preparation of $MgH_2$/Mg nanoparticles inside a matrix of reduced graphene oxide and 1,4-bis(triethoxysilyl)benzene (rGO-BTB)**

## 2. Experimental Section

### 2.1. Materials

Graphite flakes (< 20 µm), potassium chlorate (KClO$_3$, ≥ 99.0 %), dodecylamine (≥ 99.0 %), 1,4-bis(triethoxysilyl)benzene (BTB, 99.0 %), 1-butanol (anhydrous, 99.8 %), Tetrahydrofuran (anhydrous, 99.9 %), and di-n-butylmagnesium (MgBu$_2$, 1.0 M in heptane) were acquired from Sigma-Aldrich. Sulfuric acid (H$_2$SO$_4$, 95.0-97.0 %) and nitric acid (HNO$_3$, 65.0 %) were purchased from Boom BV. Milli-Q water (resistivity 18 MΩ·cm, 25 °C) was freshly produced before use. All the chemicals were used as received.

### 2.2 Synthesis of MgH$_2$/rGO-BTB nanocomposites

The layered heterostructure of reduced graphene oxide and organosilica, rGO-BTB, was prepared with a soft-template method[40] and the experimental details are described in the Supporting Information.

The MgH$_2$/rGO-BTB nanocomposites were prepared by the so-called bottom-up method[41] as indicated in Scheme 1. Sample handling and storage were conducted under an inert atmosphere in an argon-filled glovebox (Lab 2000, Etelua Intertgas System Co., Ltd.) with a gas-circulation system. Before being brought in contact with MgBu$_2$, rGO-BTB was vacuum dried at 180 °C for 3 h to eliminate all moisture from the porous structure. rGO-BTB was impregnated with MgBu$_2$ with intermittent vacuum evaporation of the excess solvent (heptane) to absorb the maximum amount. MgBu$_2$/rGO-BTB was transferred from the airtight flask to a cylindrical quartz container and sealed in an autoclave. The precursors in the autoclave were hydrogenated at 55 bar H$_2$ pressure at 180 °C for 10 h following the reaction MgBu$_2$ + H$_2$ → MgH$_2$ + 2C$_4$H$_{10}$↑. Two batches were synthesized with respectively 10 wt.% and 20 wt.% Mg loading of the nanocomposites, denoted as MgH$_2$/rGO-BTB-10 and MgH$_2$/rGO-BTB-20.

### 2.3 Thermal desorption and regeneration

The hydrogen desorption properties of MgH$_2$/rGO-BTB were studied by temperature programmed desorption (TPD) using a Micromeritics AutoChem II 2920 apparatus equipped with a TCD detector in conjunction with mass spectrometry. The MgH$_2$/rGO-BTB composites were loaded without air exposure, and the H$_2$ partial pressure and TCD signal were recorded while applying a constant temperature rate of 5 °C min$^{-1}$ from 25 to 500 °C with 25 mL min$^{-1}$ Ar flow.

Then we performed rehydrogeneration/dehydrogenation experiments where the starting material was cycled at different temperatures to check for recyclability. Here we report in detail on the experiments where Mg/rGO-BTB was exposed in the autoclave to 12 bar H$_2$ pressure at 180 °C for 18 h to obtain MgH$_2$/rGO-BTB and then dehydrogenated in the TPD setup at 200 °C; four cycles were performed.

**2.4 Materials characterization**

Fourier-transform infrared spectroscopy (FTIR) spectra (resolution of 4 cm$^{-1}$, 32 scans) were acquired using a Bruker Vertex 70 spectrometer in the 500-4000 cm$^{-1}$ range. X-ray diffraction (XRD) spectra were collected using a D8 Advance Bruker diffractometer with Cu K$_\alpha$ radiation ($\lambda$=1.5418 Å) employing a 0.25° divergent slit and a 0.125° anti-scattering slit; the patterns were recorded in the 2θ range from 2° to 80°, with steps of 0.02° and a counting time of 2 s per step. With the air sensitive samples (MgH$_2$/rGO-BTB), a dome-shaped airtight sample holder was used to prevent the reaction with air during the measurement, and the loading of the sample in the holder was performed in an argon-filled glove box,[42] and the patterns were recorded in the 2θ range from 10° to 80° with a step of 0.017° and a time step of 60 s for a total of 30 min per scan. The nitrogen adsorption-desorption isotherms were measured at -196 °C on a Micromeritics ASAP 2420 V2.05 porosimeter. All samples were degassed overnight at 120 °C under vacuum before analysis. The specific surface area was evaluated with the Brunauer-Emmett-Teller (BET) model by fitting the N$_2$ desorption isotherm; the pore volume was determined at $P/P_0$ = 0.995 and the pore size distribution was analyzed by the non-local density functional theory (NLDFT) method.[43] Scanning electron microscopy (SEM) micrographs were collected with a FEI Philips FEG-XL30s microscope; the morphology was characterized using a JEOL 2010 operating at 200 kV.

A probe and image Cs aberration corrected 30-300 kV Thermo Fisher Scientific Thenis Z (scanning) transmission electron microscope (S/TEM) equipped with the dual X-ray detector was employed for the structural and $MgH_2$ particle size characterization. The acceleration voltage was set at 300 kV. Images were acquired using high-angle annular dark-field (HAADF)-STEM (21 mrad convergence semi-angle, 50 pA probe current, 31-186 mrad collection angles of the HAADF detector), bright-field TEM and dark-field TEM. Elemental maps were acquired using energy dispersive X-ray spectroscopy (EDS)-STEM. X-ray photoelectron spectroscopy (XPS) data of rGO-BTB were collected with a SSX-100 (Surface Science Instruments) spectrometer equipped with a monochromatic Al $K_α$ X-ray source (hv =1486.6 eV). The measurement chamber pressure was maintained at $1×10^{-9}$ mbar during data acquisition; the photoelectron take-off angle was 37° with respect to the surface normal. The diameter of the analyzed area was 1000 μm and the energy resolution was 1.26 eV. For the XPS measurement rGO-BTB was dispersed in chloroform and drop-casted on a thin gold film, grown on mica.[44] *In-situ* XPS spectra of $MgH_2$/rGO-BTB-10 were collected with a high-resolution Scienta spectrometer employing a monochromatic Al $K_α$ X-ray source and equipped with a hemispherical electron analyzer (Scienta R4000). The spectra of wide scans and core level regions were acquired at a base pressure of $9×10^{-10}$ mbar, and the overall energy resolution was 0.35 eV. $MgH_2$/rGO-BTB-10 sample was dispersed in anhydrous tetrahydrofuran and drop-casted on a thin gold film grown on mica[44] in an argon-filled glove box, and then transferred to the load-lock chamber under Ar atmosphere. Spectra were acquired after annealing at 200 and 300 °C for 2 h. XPS spectral analysis included a Shirley background subtraction and fitting with a minimum number of peaks consistent with the expected composition of the probed volume, taking into account the experimental resolution. Peak profiles were taken as a convolution of Gaussian and Lorentzian functions; with the help of the least squares curve-fitting program WinSpec (LISE, University of Namur, Belgium). Binding energies (BEs) were referenced to Au4$f_{7/2}$ photoemission peak[45] centered at a binding energy of 84.0 eV and are accurate to ±0.1 eV when deduced from the fitting procedure. All measurements were carried out on freshly prepared samples; three different spots were measured on each sample to check for reproducibility.

### 3. Results and Discussions

### 3.1 Chemical and morphological characterizations of the rGO-BTB heterostructure matrixes

It is important to verify whether or not the composition, structure and morphology of rGO-BTB heterostructure agree with those reported in previous studies.[40] To this end, at each stage of the synthesis, the materials were characterized by a variety of techniques, firstly by FTIR to confirm the successful incorporation of the dodecylamine and the silica precursor BTB in the interlayer space, as shown in Figure S1(a). Compared to the spectra of GO, the spectrum of dodecylamine-intercalated GO shows two additional peaks at 2847 cm$^{-1}$ and 2919 cm$^{-1}$ that are ascribed to the asymmetric and symmetric stretching vibrations of C-H bonds.[46] Another additional peak located at 1560 cm$^{-1}$ is due to the N-H vibrations of the dodecylamine and proves that dodecylamine is present in the layered structure. A confirmation of the presence of organosilica matrix in the composite comes from the bands centered at 520 cm$^{-1}$, 1065 cm$^{-1}$ and 1151 cm$^{-1}$, which correspond to Si-O-Si vibrations, as well as from the peaks at 690 cm$^{-1}$ and 950 cm$^{-1}$, due to the O-Si-O stretching modes.[47] After calcination, the broad peak centered at 3400 cm$^{-1}$, ascribed to the -OH stretching vibration and stretching vibration of C-H, vanishes.[48] This, together with the disappearance of the band centered at 1560 cm$^{-1}$, attributed to N-H bonds, indicates that during the pyrolysis dodecylamine and the functional groups on the surface of graphene oxide were removed.[40]

The successful intercalation of dodecylamine and BTB in the interlayer space of GO can be further confirmed by XRD, which allows to estimate the interlayer distance between the graphene oxide platelets, as shown in Figure S1 (b). By applying the Bragg equation, one can derive the basal $d_{001}$-spacing, which in pristine graphene oxide amounts to 7.6 Å, but becomes 18.5 Å after intercalation with dodecylamine.[47] This basal plane spacing corresponds to an interlayer separation $\Delta$ =18.5-6.1 Å=12.4 Å, where 6.1 Å represents the thickness of a single GO layer.[49] This value is in accordance with the chain length of dodecylamine. For GO intercalated with dodecylamine and BTB, the basal plane spacing is even larger, namely 27.0 Å, and the corresponding interlayer separation $\Delta$=27.0-6.1 Å=20.9 Å. This points to successful further expansion of the interlayer space and suggests the formation of a silica network from BTB. For the heterostructure rGO-BTB, no sharp 001 diffraction at lower angles (2-10°) but a very broad peak can be observed. This indicates that the graphene layers are no longer stacked but exfoliated in thin platelets of very few layers due to the violent expansion upon heating rGO-BTB.

To verify the chemical integrity as well as the types of chemical bonds in rGO-BTB, X-ray photoelectron spectroscopy (XPS) was employed. The overview spectrum attests to the presence of all the expected elements (Figure S2 (a)). The spectrum of the C1$s$ core level region of rGO-BTB, shown in Figure S2(b), shows four contributions: the spectral signature of the C=C bonds of graphene is centered at a BE of 284.8 eV, and makes up 70.4 % of the total C1$s$ intensity. The contribution due to C-OH bonds, at a BE of 285.9 eV, represents 21.2 % of the total C1$s$ intensity, while the contributions at BEs of 287.2 eV and 289.1 eV are respectively ascribed to the C=O and C(O)O bonds. The presence of Si-O-C bonds of BTB grafted to the oxygen-containing groups of the graphene oxide surface or to Si-O-Si bonds resulting from a sol-gel reaction between BTB molecules is supported by the Si2*p* and Si2*s* core level peaks in Figure S2(c) and (d). Taken together, the XPS spectra confirm that the reduced graphene oxide layers were successfully pillared with the silica precursor BTB by the sol-gel reaction.

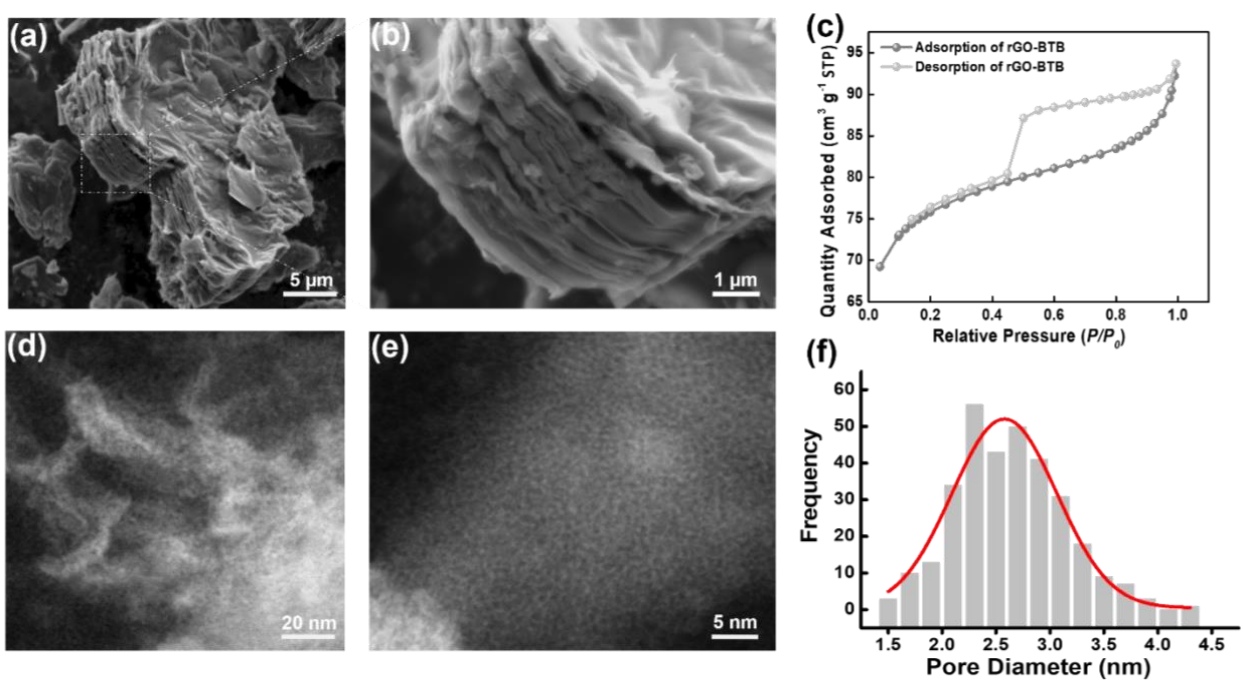

**Figure 1 Characterization of rGO-BTB: (a) and (b) SEM images; (c) N$_2$ adsorption-desorption isotherm; (d) HAADF-STEM image; (e) HAADF-STEM image zoomed in from (d) and (f) pore size distribution calculated from (e).**

The morphology of the rGO-BTB composite was initially determined via scanning electron microscopy (SEM). As shown in Figure 1(a), well-assembled multilayer platelets with smooth surfaces can be observed, and the sample consists of many such thin flakes owing to the partial exfoliation during the calcination process. The enlarged sectional view (Figure 1(b)) further confirms that mesoporous organosilica must be dispersed in between layers leading to a very regular pillared structure.

Nitrogen adsorption-desorption measurements were performed to determine the surface area and to provide information on the pore structure. A typical isotherm is presented in Figure 1(c). The sudden release of $N_2$ at $P/P_0 \approx 0.5$ gives rise to a type H4 hysteresis loop, commonly ascribed to slit-shaped pores in layered materials based on the IUPAC classification.[50] The adsorption branch of the isotherm reveals a type II plateau, and at low relative pressures, $N_2$ adsorption increases significantly with pressure, indicating that a significant amount of micropores/mesopores are accessible. The specific surface area (SSA) and pore volume were calculated to be 302±4 $m^2/g$ and 0.14 $cm^3/g$, respectively.

In order to further examine the structural characteristics, high-angle annular dark-field scanning transmission electron microscopy (HAADF-STEM) images of the rGO-BTB sample were collected at low and high magnification and typical examples are shown in Figure 1(d) and (e). In the lower magnification image (Figure 1(d)), the mesopores are distributed homogeneously in the material, while in the higher magnification image (Figure 1(e)) the nanoporous structure can be more clearly observed. The pore width distribution, determined by calculating the size of all pores visible in this image with the help of the Image J software, is shown in Figure 1(f) and peaks at a pore size of 2.5 nm.

## 3.2 Evaluation of the magnesium hydride filled mesoporous heterostructure matrixes

Once sure that the desired porous structure was achieved, we proceeded with the growth of $MgH_2$ nanoparticles in the rGO-BTB mesoporous matrix, as shown in the representation in Scheme 1, by first impregnating the latter with $MgBu_2$ and then hydrogenating (as described above in the Experimental section) for 10 h. Two batches, denoted with $MgH_2$/rGO-BTB-10 and $MgH_2$/rGO-BTB-20, were prepared, the first with 10 wt.% of $MgH_2$ introduced into the rGO-BTB heterostructure and the second one with 20 wt.%.

The X-ray diffraction patterns of pristine rGO-BTB and of rGO-BTB with the two different magnesium hydride loadings are shown in Figure 2(a). For the pattern of MgH$_2$/rGO-BTB-20, the most intense diffraction peaks originate from the MgH$_2$ tetragonal phase.[25] The Scherrer equation[51] gives a particle size of 27 nm for MgH$_2$. Since the rGO-BTB matrix has a narrow pore size distribution around 2.5 nm, such big particles likely grew on the outside surface of rGO-BTB flakes, unless the layers locally ruptured forming a small cavity.[23] No diffraction peaks from MgH$_2$ can be noticed for MgH$_2$/rGO-BTB-10, suggesting that no large particles grew outside the matrix in this case and that the MgH$_2$ nanoparticles, which were formed, have coherence lengths that are too small to diffract the X-rays.[25]

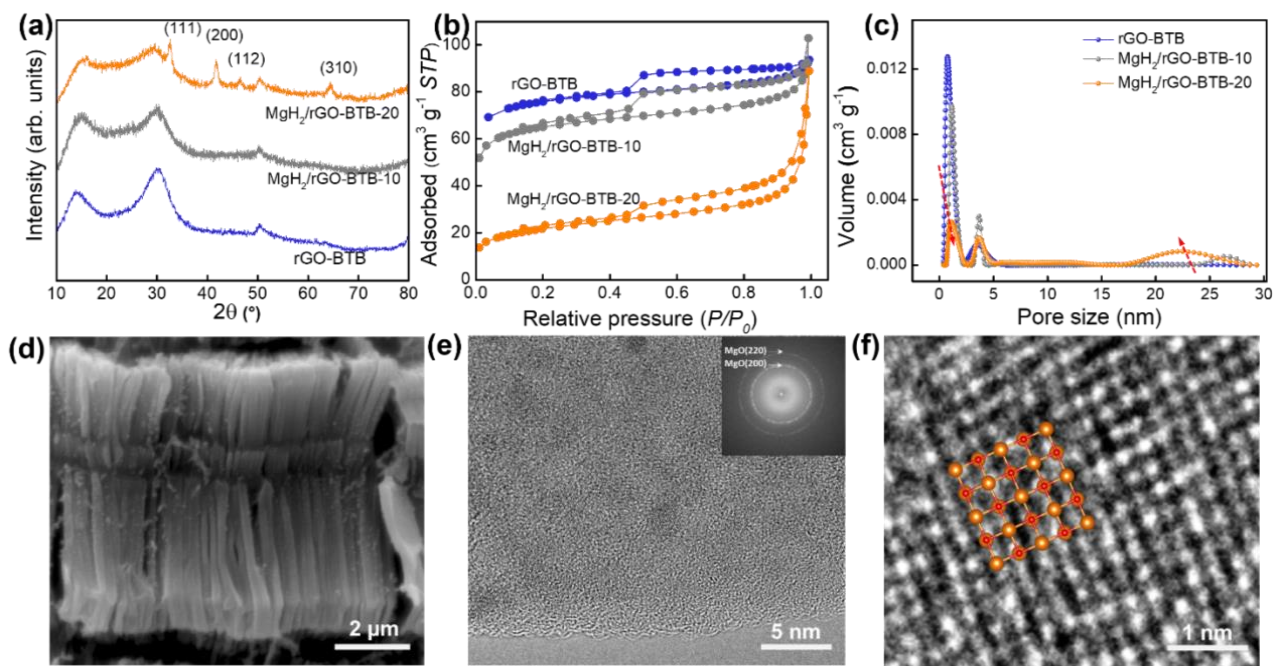

**Figure 2.** (a) X-ray diffraction patterns, (b) N$_2$ adsorption-desorption isotherms and (c) NLDFT pore size distribution of rGO-BTB, MgH$_2$/rGO-BTB-10 and MgH$_2$/rGO-BTB-20. (d) SEM image of MgH$_2$/rGO-BTB-10; (e) TEM image of MgH$_2$/rGO-BTB-10, the insert shows the corresponding fast Fourier transform pattern; (f) HRTEM image of MgH$_2$/rGO-BTB-10, overlay: MgO atomic structure.

In order to gain further insight into the porous structure after the formation of $MgH_2$, nitrogen adsorption-desorption measurements were performed and the corresponding isotherms are shown in Figure 2(b). The specific surface area and pore volumes were calculated and are listed in Table S1. The BET specific surface areas and pore volumes are found to be smaller than for pristine rGO-BTB and decrease significantly with increasing amounts of $MgH_2$. This is mainly due to the blocking of the pores in the heterostructure by $MgH_2$ particles, which can be further confirmed by the NLDFT pore sized distribution evolutions in Figure 2(c).[25] For $MgH_2$/rGO-BTB-10, the change of the isotherm was limited, testifying to a decrease of 38.0 % of the specific surface area and 33.8 % of the pore volume with respect to pristine rGO-BTB, which means the magnesium hydride nanoparticles do not fill all the pores. However, for $MgH_2$/rGO-BTB-20, the micropore volume decreased significantly (86.8 % as compared to pristine rGO-BTB), while the mesoporous volume increased, indicating that the $MgH_2$ particles preferentially filled the micropores rather than the mesopores. The excess larger crystals on the outside surface of the rGO-BTB heterostructure lead to an increase of the mesoporous volume in agreement with the XRD result.

These conclusions are further supported by the field-emission scanning electron microscopy (FE-SEM) and transmission electron microscopy (TEM) images shown in Figure 2(d-f) ($MgH_2$/rGO-BTB-10) and Figure S3 ($MgH_2$/rGO-BTB-20) of the Supporting Information. The SEM image (Figure 2(d)) shows that rGO-BTB in $MgH_2$/rGO-BTB-10 largely retains its original layered morphology, while hardly any agglomerated particles located on the outside of the layered structure can be observed. The bright-field TEM images (Figure 2(e), Figure 3(a)) of $MgH_2$/rGO-BTB-10 show homogenously distributed particles throughout the rGO-BTB matrix, which have an MgO crystal structure as directly observable from the atomically resolved structure of a single particle (Figure 2(f)) and the fast Fourier transform (FFT) of the entire collection of particles (inset of Figure 2(e)) revealing the (111), (200) and (220) fcc MgO planes.[52] In contrast, the larger particles have not been oxidized, but in fact remained fully hydrogenated $MgH_2$ crystals, as the characteristic d(110) interlayer space of 0.321 nm of $MgH_2$ is visible in atomically resolved images, shown in Figure S3 (a) and (b).[53] This evidence indicates that the sample transfer in air for the TEM measurements is most likely cause for the complete oxidation of small $MgH_2$ particles.[26]

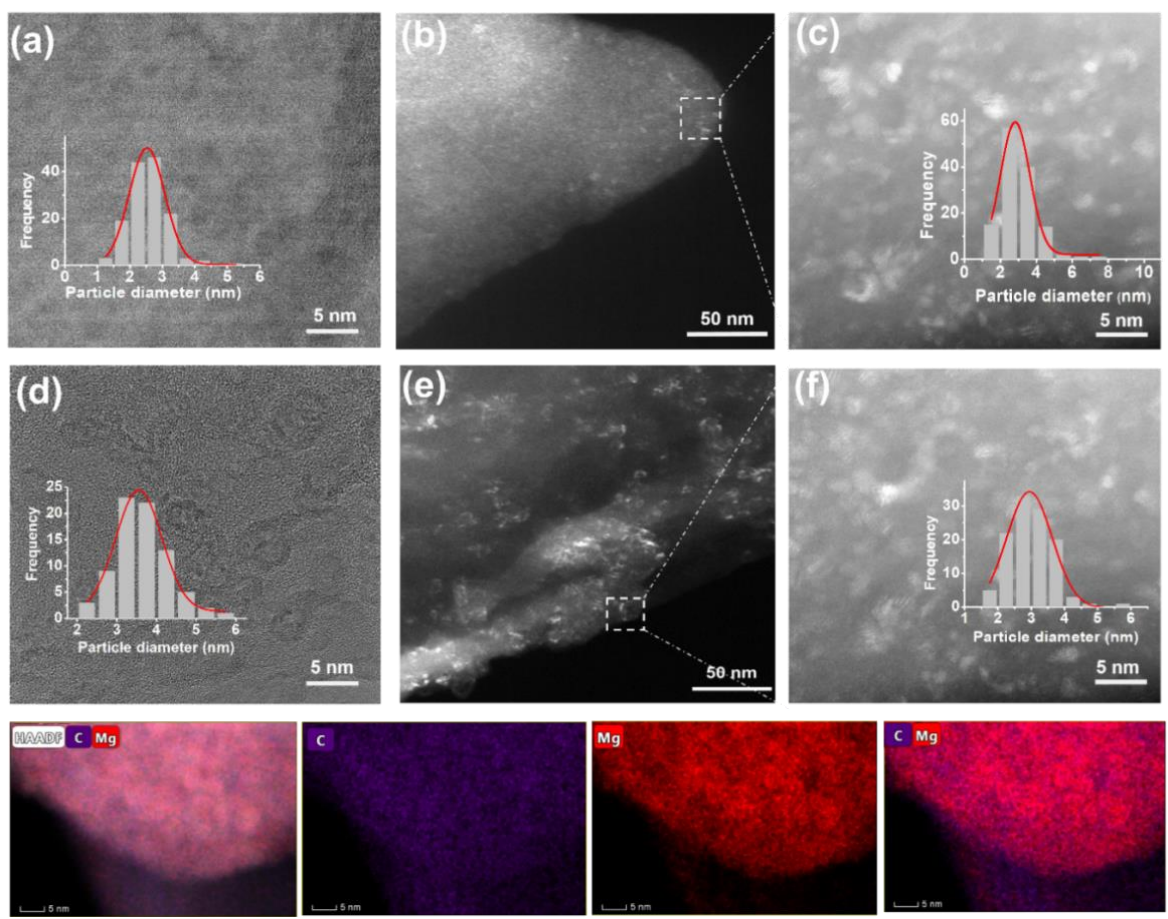

**Figure 3. (a) bright-field and (b), (c) dark-field TEM images of $MgH_2$/rGO-BTB-10, the insets of (a) and (c) show the corresponding particle size distribution of the image. (d) bright-field and (e), (f) dark-field TEM images of $MgH_2$/rGO-BTB-20, the insets of (d) and (f) show the corresponding particle size distribution of the image. The bottom panels: HAADF-STEM and EDS elemental maps for C, Mg, and C combined with Mg in $MgH_2$/rGO-BTB-10.**

Since the particle size is a significant parameter for the hydrogen storage properties of $MgH_2$ in the porous structure, the particle size distribution was extracted from the bright-field and dark-field TEM images of $MgH_2$/rGO-BTB-10 and $MgH_2$/rGO-BTB-20. In the bright-field TEM image (Figure 3(a)) of $MgH_2$/rGO-BTB-10, the particle size distribution is centered at 2.5 nm, and in the dark-field TEM of the same batch, the particle size distribution peaked at 3.0 nm. As expected from the XRD results, the TEM images of $MgH_2$/rGO-BTB-20 (Figure 3(d-f)) and $MgH_2$/rGO-BTB-20 (Figure 3(d-f) and Figure S3 (b)) show the several larger particles on the external surface of rGO-BTB. However, if one excludes these large particles from the calculation, the particle size distributions, shown in the insets of Figure 3(d) and 3(f), were again centered on 2.5 nm, as for $MgH_2$/rGO-BTB-10. The SEM images of $MgH_2$/rGO-BTB-20 (Figure S3(c)) seem to point to crystal growth from the inside of the porous structure to the external surface; a large amount of $MgH_2$ crystals can be distinguished on the surface and dendrites are seen to have grown from the inside outwards. HAADF-STEM images and EDS elemental maps, as presented in the bottom panel of Figure 3, further verify the successful confinement of majority of the $MgH_2$ particles in the rGO-BTB heterostructure. Mg and C are homogeneously distributed in the material, while Mg is also dispersed in the porous structure and surrounded by C, in good agreement with the results of the other characterization techniques.

**3.3 Hydrogen desorption properties of $MgH_2$/rGO-BTB composites.**

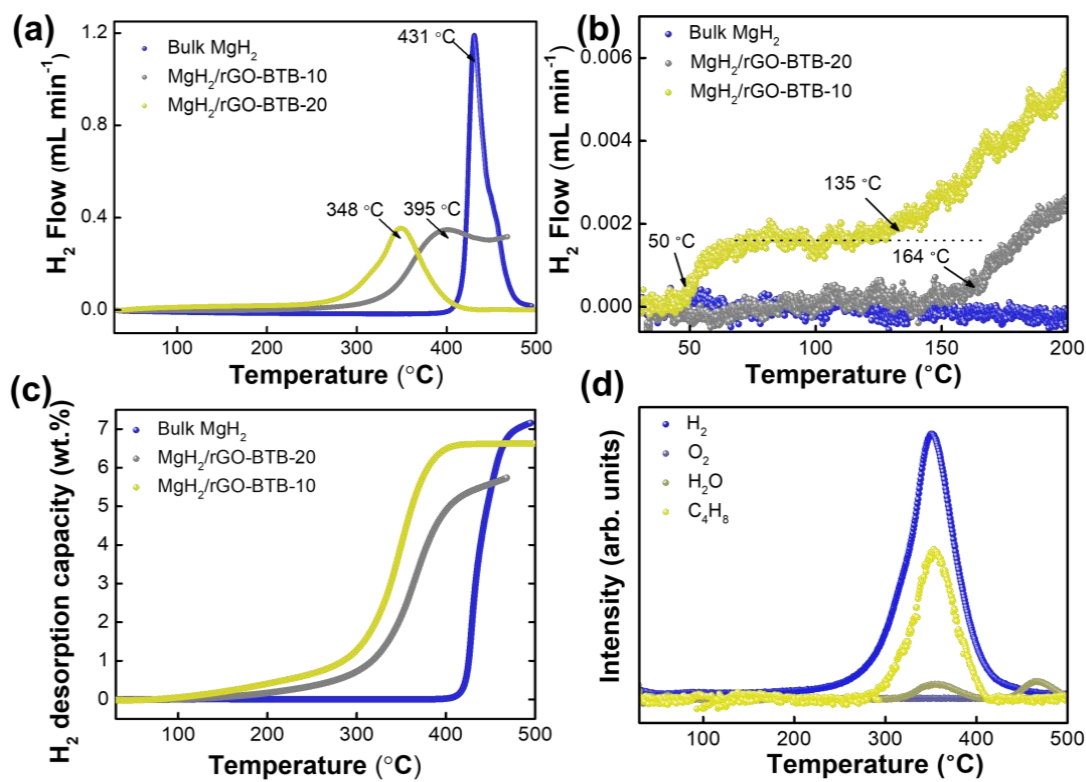

**Figure 4.** (a) Temperature programmed desorption (TPD) spectra of bulk MgH$_2$ and MgH$_2$ particles confined in the rGO-BTB structure, recorded with a heating rate of 5 °C min$^{-1}$; (b) TPD spectra of bulk MgH$_2$ and confined MgH$_2$ particles in rGO-BTB structures from 30 °C to 200 °C: enlarged view of the onset of desorption in (a); (c) hydrogen desorption capacity of bulk MgH$_2$ and MgH$_2$ particles embedded in the rGO-BTB structure from 30 °C to 500 °C - the capacity is based on the confined MgH$_2$ weight; (d) mass spectrometry analysis upon heating to 500 °C for MgH$_2$/rGO-BTB-10.

In order to explore the hydrogen storage properties of MgH$_2$ particles confined in the rGO-BTB heterostructure, temperature programmed desorption (TPD) experiments of the bulk MgH$_2$ and of the two batches with different MgH$_2$ loading were carried out; the results are shown in Figure 4(a). One observes that a temperature as high as 431 °C is needed for bulk MgH$_2$ to release H$_2$; such a high dehydrogenation temperature and sluggish kinetics are far from meeting the requirements for automotive applications. In contrast, MgH$_2$ particles confined in the mesoporous heterostructure show superior dehydrogenation properties:

MgH$_2$/rGO-BTB-20 exhibits an onset temperature of desorption around 160 °C, 271 °C lower than the temperature threshold of bulk MgH$_2$. Remarkably, for the MgH$_2$/rGO-BTB-10, an onset temperature of 50 °C can be observed (Figure 4(b)), an impressive 381 °C lower value than that of bulk MgH$_2$. This low onset is also evident in the detailed TDS spectrum collected in the temperature region of 40-100 °C and shown in Figure S4. Figure 4(b) shows a plateau between 50-135 °C, while exhibiting an exponential growth after 135 °C, which is close to the onset temperature of MgH$_2$/rGO-BTB-20. That implies that the onset temperature for hydrogen release at 50 °C could be physisorbed H$_2$ in the unfilled mesopores, while 135 °C could be the temperature for the nanosized MgH$_2$ particles to start releasing hydrogen. The reason why the desorption of physisorbed H$_2$ is not visible for MgH$_2$/rGO-BTB-20, is clear from the BET results, which indicate that the mesoporous structure is largely blocked by the MgH$_2$ particles.

The superior dehydrogenation properties of MgH$_2$/rGO-BTB can be ascribed to the nanometer sized MgH$_2$ particles in the mesoporous heterostructure. Nanometer sized MgH$_2$ is thermodynamically less stable, resulting in lower activation energies for desorption, and therefore, a lower onset temperature than bulk MgH$_2$. In addition, nanometer sized MgH$_2$ crystals boost the kinetics of dehydrogenation because of the intrinsically short diffusion paths for hydrogen, which represent the rate-limiting step for the hydrogenation or dehydrogenation processes. The catalytic properties of the porous scaffold might also play a role in the improvement of the kinetics.[54] The unsaturated carbon atoms from reduced graphene oxide bind with MgH$_2$ located between the layers, resulting in electron transfer from Mg to rGO and thus weaker Mg-H bonds, a scenario in which hydrogen may be released with smaller activation energy.[55] Furthermore, while the TPD desorption isotherm peak of MgH$_2$/rGO-BTB-10 is symmetric, the peak of MgH$_2$/rGO-BTB-20 is broader and asymmetric, pointing to two populations of nanoparticles with different desorption kinetics. The small shoulder at higher temperature close to the desorption temperature of bulk MgH$_2$ is attributed to the larger particles outside the rGO-BTB matrix. This TPD profile is hence consistent with the bimodal particle size distribution shown by the TEM study (Figure 3(d)). The desorption is not completed at 470 °C: desorption from MgH$_2$ located outside of the matrix and the ordered mesoporous heterostructure continues at higher temperature.[56] However, the temperature for maximum hydrogen desorption of 348 °C for the batch of MgH$_2$-rGO-BTB-10 is higher than what previous work[16] found for particle sizes <10 nm. This indicates that not only the particle size influences the maximum desorption temperature, but also the interaction between the MgH$_2$ particles and the rGO matrix as well as the pillaring structures may affect hydrogen desorption properties of MgH$_2$. In future work, different pillars will be used to investigate this observation.

The dehydrogenation capacities of confined MgH$_2$ and the bulk MgH$_2$ are presented in Figure 4(c). It should be noted that the desorption capacity of the MgH$_2$/rGO-BTB was obtained by excluding the weight percentage of the heterostructure matrixes to compare the hydrogenation properties of the confined Mg nanoparticles. The desorption capacity of bulk MgH$_2$ reaches 7.1 wt.%; in contrast, the confined MgH$_2$ in the heterostructure shows relatively lower desorption capacity. MgH$_2$/rGO-BTB-10 has around 6.6 wt.% of reversible H$_2$ storage capacity related to confined MgH$_2$, while for MgH$_2$/rGO-BTB-20, the desorption capacity reaches only 5.7 wt.%, presumably because the larger MgH$_2$ crystals located outside of the matrix become partially oxidized or reacted with residual oxygen-containing groups on the external surface of the matrix.[57]

We collected mass spectra during TPD to monitor the chemical species released during the MgH$_2$/rGO-BTB-10 dehydrogenation process; the results are shown in Figure 4(d). As expected, the hydrogen signal is the largest one; the release of butane was expected because it is the byproduct of the hydride elimination reaction that transforms the MgBu$_2$ solution into magnesium hydride.[24] There is no oxygen signal detected, which means no $O_2$ is released from the heterostructure and no $O_2$ contamination is evident. However, desorption peaks related to water were observed at similar temperatures to those of hydrogen, namely centered at 350 °C and 475 °C; this might point to a water-producing reaction involving the decomposition of $Mg(OH)_2$ or the reaction between the dehydrogenated Mg and oxygen-containing groups on the porous matrix.

**3.4 Hydrogen storage reversibility of MgH$_2$/rGO-BTB-10 composite**

The reversibility of the performance of hydrogen storage materials is considered as an important requisite for automotive application.[58] Mesoporous silica has been proposed as a promising nanoscale template for metal hydrides, such as $NaBH_4$,[59] $NaAlH_4$,[60] and $LiBH_4$[61] by melt infiltration, however the reaction between the silanol groups in the pore channel and metal hydrides at high temperature resulted in an irreversibility loss of hydrogen storage capacity.[62] For the MgH$_2$/Mg hydrogen storage system, Zhu *et al.*[63] have reported that MgH$_2$ confined in mesoporous silica shows excellent reversibility at 250 °C. In order to investigate the reversibility of MgH$_2$/rGO-BTB-10, the isothermal hydrogen absorption/desorption cycling measurements were carried out at 200 °C, as illustrated in Figure 5(a). MgH$_2$/rGO-BTB-10 at 200 °C has a hydrogen storage capacity of 1.83 wt.% during the first cycle and in the following three cycles, a high reversible capacity of 1.62 wt.% was achieved, in other words, 88.5 % of the original capacity was preserved (Figure 5(b)). It is worth noting that the hydrogen was released within 20 min, indicating that the desorption kinetics did not degrade in these four cycles, indirectly confirming that the confinement of the nanoparticles in the heterostructure inhibits aggregation.[64] However, when dehydrogenation is performed at temperatures higher than 250 °C, MgH$_2$/rGO-BTB-10 is no longer stable and reversible hydrogen absorption/desorption is no more possible.

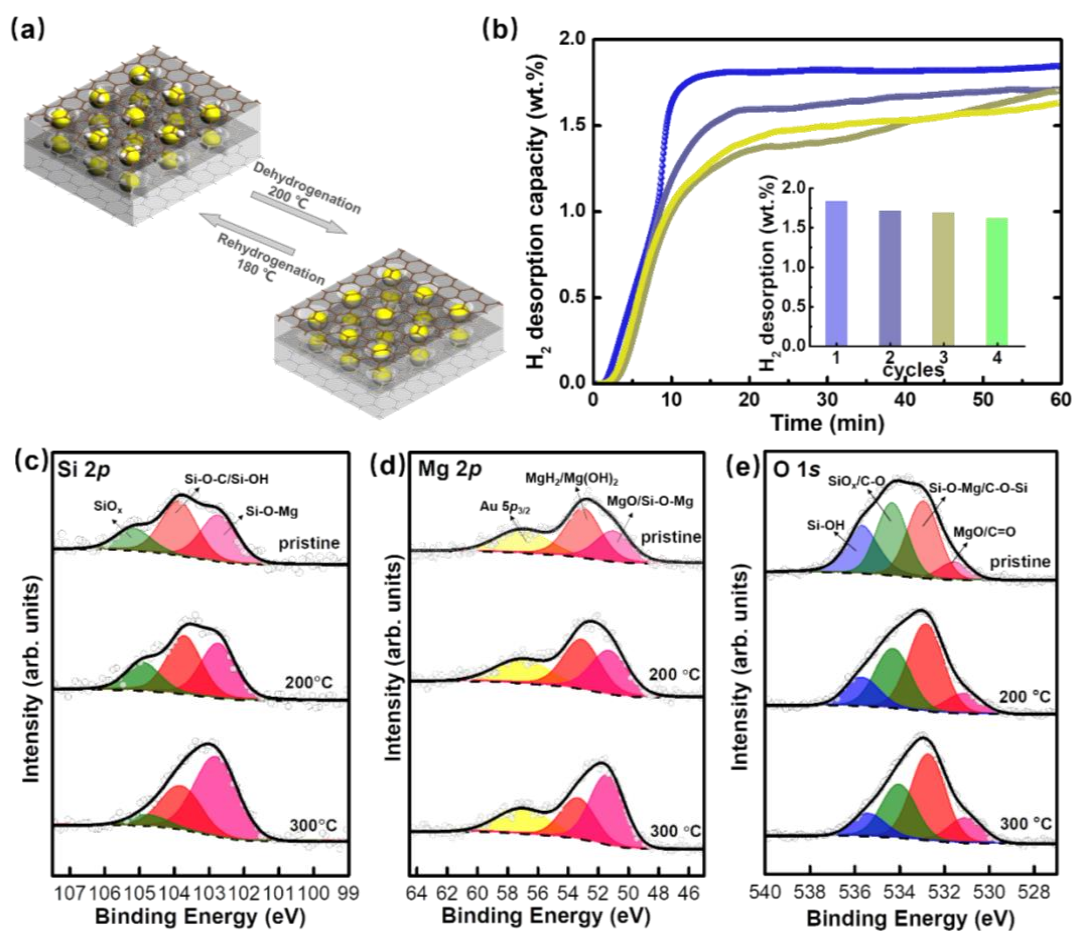

**Figure 5.** (a) Schematic illustration of the rehydrogenation and dehydrogenation process of Mg/rGO-BTB-10 and MgH$_2$/rGO-BTB-10 composites; (b) hydrogen desorption of reversibility measurement for MgH$_2$/rGO-BTB-10 at 200 °C (absorption at 180 °C; desorption at 200 °C), inset: the hydrogen desorption capacity value of each cycle; *in situ* X-ray photoelectron spectroscopy of the (c) Si2*p*, (d) Mg2*p* and (e) O1*s* core level regions of MgH$_2$/rGO-BTB-10 after heating to different temperatures as indicated.

In order to gain insight into the chemical changes of MgH$_2$/rGO-BTB-10 with increasing temperature, we employed *in-situ* X-ray photoelectron spectroscopy. The spectra of the Si2*p* core level region are presented in Figure 5(c); for pristine MgH$_2$/rGO-BTB-10 three contributions are needed in the fit: the spectral signature of Si-O-C overlaps with Si-OH centered at the BE of 103.6 eV, while the SiO$_x$ component can be observed centered at 105.2 eV.[65] The presence of Si-O-Mg bonds is confirmed by the component peaked at a BE of 102.6 eV.[66] After heating at 200 °C and consequent dehydrogenation, slight changes in the relative intensities of the three components in Si2$p$ core level spectrum can be observed; instead after dehydrogenation at 300 °C, the relative intensities of the components vary more dramatically: the intensity of the signal due to Si-O-Mg increases, while that of the components originating from C-O-Si/Si-OH and SiO$_x$ decreases, implying that the reaction between dehydrogenated Mg and the silanol groups could be responsible for the irreversibility of hydrogen storage.

The spectrum of the Mg2$p$ core level region of MgH$_2$/rGO-BTB-10 is shown in Figure 5(d). It partially overlaps with the Au5$p_{3/2}$ line of the substrate, peaked at 57.0 eV,[45] and can be fitted with two contributions centered at BEs of 50.9 eV and 52.8 eV, respectively. The latter contains the signals of MgH$_2$ and Mg(OH)$_2$, while the former is ascribed to Si-O-Mg/MgO.[67]

The O1$s$ core level region of MgH$_2$/rGO-BTB-10 is presented in Figure 5(e), and contains four contributions: a first peak at a BE of 531.5 eV that derives from MgO/C=O;[68] a second one at a BE of 532.9 eV, which corresponds to the relatively electron-poor oxygen from Mg-O-Si/C-O-Si bonds, and two contributions centered at 534.2 eV and 535.6 eV, stemming from SiO$_x$/C-O and Si-OH, respectively.[69]

After heating to 200 °C a slight decrease of the intensity of MgH$_2$/Mg(OH)$_2$ component of the Mg2$p$ core level can be observed, while that of the Mg(OH)$_2$ component remains stable at this temperature,[70] indicating the successful dehydrogenation of MgH$_2$. The concomitant small intensity increase of the MgO/Si-O-Mg component could explain why the hydrogen storage capacity is slightly lower in the consecutive cycles. However, after dehydrogenation at 300 °C, the narrower full width at half maximum (FWHM) of the Mg2$p$ and O1$s$ core level line signals more important chemical changes. A significant decrease of the spectral intensity due to MgH$_2$/Mg(OH)$_2$, and an important increase of that corresponding to MgO/Mg-O-Si can be observed in the Mg$2p$ core level region, and indicate that the dehydrogenation of MgH$_2$ is accompanied by the decomposition of Mg(OH)$_2$ to give rise to Si-O-Mg bonds or MgO. In addition, the significant increase of the fingerprint of Si-O-Mg bonds in both Si$2p$ and O$1s$ core level photoemission signals, while the MgO/C=O intensity in O$1s$ core level hardly changes, imply that dehydrogenated Mg tends to react with Si-OH to form Si-O-Mg bonds instead of MgO.

Further spectral evidence confirms this picture: the photoemission spectra of the C$1s$ core level region of MgH$_2$/rGO-BTB-10 are shown in Figure S5. Since rGO-BTB is produced by calcination at 370 °C for 2 h, most of the C=O/O-C=O groups were removed in that synthesis step and the relatively thermostable bonds C-O/C-O-Si bonds remain in the framework; it is therefore no surprise that the intensity of the C=O/O-C=O component hardly decreases after heat treatments at 200 and 300 °C (Table S2).[71] The presence of oxidized magnesium is instead further corroborated by the shape of the Mg *KLL* Auger peak, shown in Figure S6: already in the pristine sample there is not only the contributions at 310.2 eV deriving from Mg(OH)$_2$ and MgH$_2$ but also the spectral signature of MgO at 308.0 eV. The Mg *KLL* Auger line therefore points to a partial oxidation of the confined MgH$_2$ particles after hydrogenation and sample transport, despite all the precautions taken (described in the section on experimental details).[68] More importantly, the MgO component becomes dominant after heating to 300 °C, while the Mg(OH)$_2$ and MgH$_2$ strongly decrease. Both the Mg$2p$ photoemission spectrum and the Mg *KLL* Auger line give evidence for an increasing oxidation of magnesium when heated to 300 °C, confirming the conclusion that the reaction between dehydrogenated Mg and the silanol groups degrades the recyclability at this high temperature. Based on the minor chemical changes at 200 °C, the reversible hydrogen storage of MgH$_2$/rGO-BTB-10 when not surpassing that temperature in the dehydrogenation could indicate that temperatures of 250 °C or higher are needed to activate the reaction between Si-OH and Mg. These findings suggest that calcination at higher temperature to remove the silanol groups and any residual oxygen-containing groups, or inactivation of the silanol groups by methylation[72] before magnesium hydride loading could be ways to realize reversible hydrogen storage at higher temperature in this type of hybrid.

## 4. Conclusions

A novel layered heterostructure of reduced graphene oxide and organosilica with high specific surface area and narrow pore size distribution was successfully synthesized by surfactant-directed sol-gel reaction of an organosilicon precursor in the interlayer space of graphene oxide. Porosimetry measurement and HAADF-STEM images revealed the mesoporous structures were with narrow pore size distribution of 2.5 nm, enabling the confinement of nanosized $MgH_2$ nanoparticles *via* wet impregnation with $MgBu_2$ followed by thermal hydrogenation. Due to the well-defined porous structure in the matrix, $MgH_2$ crystals were homogeneously distributed with the particle size of ~2.5 nm. The onset of the hydrogen desorption is observed at temperatures as low as 50 °C, and exhibits maximum desorption at 348 °C. In addition, the cycling experiments show that efficient reversible 1.62 wt.% hydrogen storage can be realized at 200 °C, indicating that the scaffold can efficiently inhibit nanoparticle aggregation and coalescence. At higher temperature, however, the hydrogen storage capacity is lost due to the gradual irreversible reaction of the Mg with the residual silanol groups. This $MgH_2$ nanoparticle confinement approach has promising prospects, and is compatible with the incorporation of catalytic or reactive additives in the porous matrix to further tune the thermodynamic and kinetic performance of the $MgH_2$ nanoparticles for hydrogen storage purposes.

**Supporting Information**

Supporting Information is available from the Wiley Online Library or from the author.

**Acknowledgments**

F.Y. acknowledges the China Scholarship Council (CSC 201704910930) and the University of Groningen for supporting his PhD studies. P.N. and P.d.J. acknowledge Dutch Research Council (NWO), ECHO grant 712. 015. 005. We would like to thank Jacob Baas for the help of XRD measurement, and Léon Rohrbach for the $N_2$ adsorption-desorption experiment. This work was supported by the Advanced Materials research program of the Zernike National Research Centre under the Bonus Incentive Scheme of the Dutch Ministry for Education, Culture and Science. This research was also co-financed by Greece and the European Union (European Social Fund- ESF) through the Operational Programme «Human Resources Development, Education and Lifelong Learning» in the context of the project "Reinforcement of Postdoctoral Researchers - 2nd Cycle" (MIS-5033021), implemented by the State Scholarships Foundation (IKY).

**Author Contributions**

P. Rudolf, P. de Jongh and D. Gournis conceived the study and finalized the manuscript. F. Yan conducted the experiments, the data analysis, and wrote the draft of the manuscript. P. Ngene performed the hydrogen storage experiments and their interpretation. E. M. Alfonsín, K. Spyrou and E. Thomou contributed to the material synthesis and characterization. S. de Graaf and B. J. Kooi performed the S/TEM characterization. O. De Luca collaborated in the XPS measurements. H.T. Cao, L.Q. Lu and Y.T. Pei participated in the SEM experiments. All authors discussed the results and commented on the manuscript.

**Conflict of Interest**

The authors declare no conflict of interest.